\documentclass[journal=aelccp,manuscript=letter,layout=onecolumn]{achemso}
\usepackage{mwe}
\usepackage{comment}
\usepackage{natbib}
\usepackage{caption}
\usepackage{subcaption}
\usepackage{multirow}
\usepackage{amsmath}
\usepackage{glossaries}
\usepackage[pdfborder={0 0 0}]{hyperref}
\hypersetup{colorlinks={true},urlcolor=blue,citecolor=blue,runcolor=blue,linkcolor=blue}
\usepackage[version=4]{mhchem}

\newacronym{oer}{OER}{oxygen evolution reaction}
\newacronym{dft}{DFT}{density functional theory}
\newacronym{ctl}{CTL}{charge transition level}
\newacronym{che}{CHE}{computational hydrogen electrode}
\newacronym{cbm}{CBM}{conduction band minimum}
\newacronym{vbm}{VBM}{valence band maximum}
\newacronym{pds}{PDS}{potential-determining step}
\newacronym{pec}{PEC}{photoelectrochemical}

\ExplSyntaxOn
\keys_define:nn { mhchem }
 {
  arrow-min-length .code:n =
   \cs_set:Npn \__mhchem_arrow_options_minLength:n { {#1} } 
 }
\ExplSyntaxOff

\title{Spontaneous Oxygen Vacancy Ionization Enhances Water Oxidation on BiVO$_4$}
\author{Nicklas Österbacka}
\affiliation{Department of Physics, Chalmers University of Technology, SE-412 96 Gothenburg, Sweden}
\author{Hassan Ouhbi}
\affiliation{Department of Physics, Chalmers University of Technology, SE-412 96 Gothenburg, Sweden}
\author{Francesco Ambrosio}
\affiliation{Dipartimento di Chimica e Biologia Adolfo Zambelli, Università di Salerno, Via Giovanni Paolo II, I-84084 Fisciano (SA), Italy}
\affiliation{Dipartimento di Scienze, Università degli Studi della Basilicata, Viale dell’Ateneo Lucano, 10-85100 Potenza, Italy}
\author{Julia Wiktor}
\affiliation{Department of Physics, Chalmers University of Technology, SE-412 96 Gothenburg, Sweden}
\email{julia.wiktor@chalmers.se}

\begin{document}

\begin{abstract}
    The influence of surface oxygen vacancies on the oxygen evolution reaction on bismuth vanadate is studied using hybrid density functional theory. Our findings reveal the thermodynamic instability of the neutral unionized defect (\ce{V_O^0}), leading to spontaneous ionization into \ce{V_O^{2+}}. By investigating the oxygen evolution reaction mechanism on both stoichiometric and oxygen-deficient surfaces, we find that surface oxygen vacancies reduce the reaction's thermodynamic overpotential, but only when the defects are ionized. Moreover, the reaction pathway involves the formation of surface-bound peroxide and superoxide ions as intermediates. Our work provides insight into the nature of surface oxygen vacancies and shines new light on how they may enhance the photoelectrochemical properties of semiconductors.
\end{abstract}

\section{Introduction}

Solar water splitting offers an attractive pathway towards clean energy, enabling the production of hydrogen gas from nothing but water and solar light. The process consists of two halves: the cathodic hydrogen evolution reaction and the anodic \gls{oer}. The latter of the two involves four proton-coupled electron transfers and thus requires a photoanode with excellent charge transfer properties to overcome its sluggish kinetics.~\cite{songReviewFundamentalsDesigning2020} While a great deal of research has been put into improving the anode, the \gls{oer} remains the bottleneck in the overall water splitting process.~\cite{fabbriOxygenEvolutionReaction2018, wangIdentifyingKeyObstacle2018, songReviewFundamentalsDesigning2020}

The ideal \gls{oer} (photo)catalyst should be cheap, stable and exhibit a moderate bandgap that allows for light absorption in the visible regime to maximize efficiency.~\cite{jiangPhotoelectrochemicalDevicesSolar2017, songReviewFundamentalsDesigning2020} On paper, monoclinic bismuth vanadate (\ce{BiVO_4}), an $n$-type semiconductor with a bandgap of 2.4 eV, fulfills these requirements.~\cite{parkProgressBismuthVanadate2013} However, the practical \gls{pec} performance of the material is limited by high charge recombination rates, low conductivity, and slow charge transfer rates. Several strategies for improving upon these shortcomings have been developed in recent years, including cocatalyst deposition and doping.~\cite{abdiEfficientBiVO4Thin2013, zhangDopingStrategyPromote2017, palaniselvamVastlyEnhancedBiVO42017, talasilaModifiedSynthesisBiVO42020} Intrinsic defects also significantly alter the properties of the host material, with several studies suggesting that an abundance of oxygen vacancies (\ce{V_O}) improves the \gls{pec} performance of \ce{BiVO_4} through enhanced charge separation and charge transfer efficiency.~\cite{fengEnrichedSurfaceOxygen2020, wuMultilayerMonoclinicBiVO42018, jinOxygenVacanciesActivating2021, wangNewBiVO4Dual2018, wangSituFormationOxygen2020} The defect could also play a more direct role in the reaction as proposed in the works by Hermans \textit{et al.}~\cite{hermansBiVO4SurfaceReduction2019} and explored computationally by Nikačević \textit{et al.}~\cite{nikacevicInfluenceOxygenVacancies2021} A complete and coherent picture of the mechanism behind the improvement is still missing, however.

In this work, we investigate the stability of the surface \ce{V_O} in \ce{BiVO4} in different charge states and find that it spontaneously ionizes. We furthermore calculate how the thermodynamic overpotential of the oxygen evolution reaction is affected by the presence of \ce{V_O} at the surface. In the neutral charge state they have no effect on this property, while ionized vacancies (\ce{V_O^{2+}}) reduce the overpotential by 0.6 V. Additionally, the proposed reaction pathway involves formation of surface-bound perodixe and superoxide species.

\section{Methods}

Defect formation energies are computed within the framework laid out by Freysoldt and Neugebauer for charged defects at surfaces.~\cite{freysoldtFirstprinciplesCalculationsCharged2018} Total energies are evaluated with a PBE0-based density functional with 22 \% exact exchange as implemented in \texttt{CP2K}.~\cite{kuhneCP2KElectronicStructure2020} A more detailed overview of the computational methodology is given in the SI. For simplicity, we consider the tetragonal scheelite structure of \ce{BiVO_4} with cell parameters ($a=b=5.147$ Å, $c=11.726$ Å) in agreement with the experimental findings of Sleight \textit{et al.}~\cite{sleightCrystalGrowthStructure1979} as the electronic properties of this surface differ very little from those of the monoclinic scheelite phase.~\cite{leeImpactSurfaceComposition2021} $2\times2\times2$ repetitions of the ideal structure with 25 Å of vacuum in the surface normal direction are used to approximate the surface. Symmetric slabs are used when considering reaction mechanisms to cancel spurious dipole moments across the system, but asymmetric slabs are used for defect formation energies as required by the scheme of Freysoldt and Neugebauer.~\cite{freysoldtFirstprinciplesCalculationsCharged2018}

The \gls{oer} mechanism typically considered in mechanistic studies consists of four steps, each involving the transfer of a proton-electron pair:

\begin{equation}
    \begin{split}
        1.\ &\ce{$\ast{}$ + H_2O -> $\ast{}$OH + (H^+ + e^-)} \\
        2.\ &\ce{$\ast{}$OH -> $\ast{}$O + (H^+ + e^-)} \\
        3.\ &\ce{$\ast{}$O + H_2O -> $\ast{}$OOH + (H^+ + e^-)} \\
        4.\ &\ce{$\ast{}$OOH -> $\ast{}$ + (H^+ + e^-) + O_2}
    \end{split}
    \label{eq:watersplitting}
\end{equation}

Here, a lone \ce{$\ast{}$} denotes the bare surface, while $\ce{$\ast{}$X}$ denotes adsorption of \ce{X}. The Gibbs free energy difference at pH 0 of each reaction step is evaluated within the \gls{che} formalism.~\cite{norskovOriginOverpotentialOxygen2004} Details of the vibrational contributions may be found in the SI. The oxygen molecule is notoriously difficult to describe using \gls{dft}, so the free energy of the total reaction, i.e., {\ce{2H2O -> O2 + 4 (H^+ + e^-)}}, is set to the experimental value of 4.96~eV.

\section{Results}

\begin{figure}
    \centering
    \includegraphics[width=0.6\textwidth]{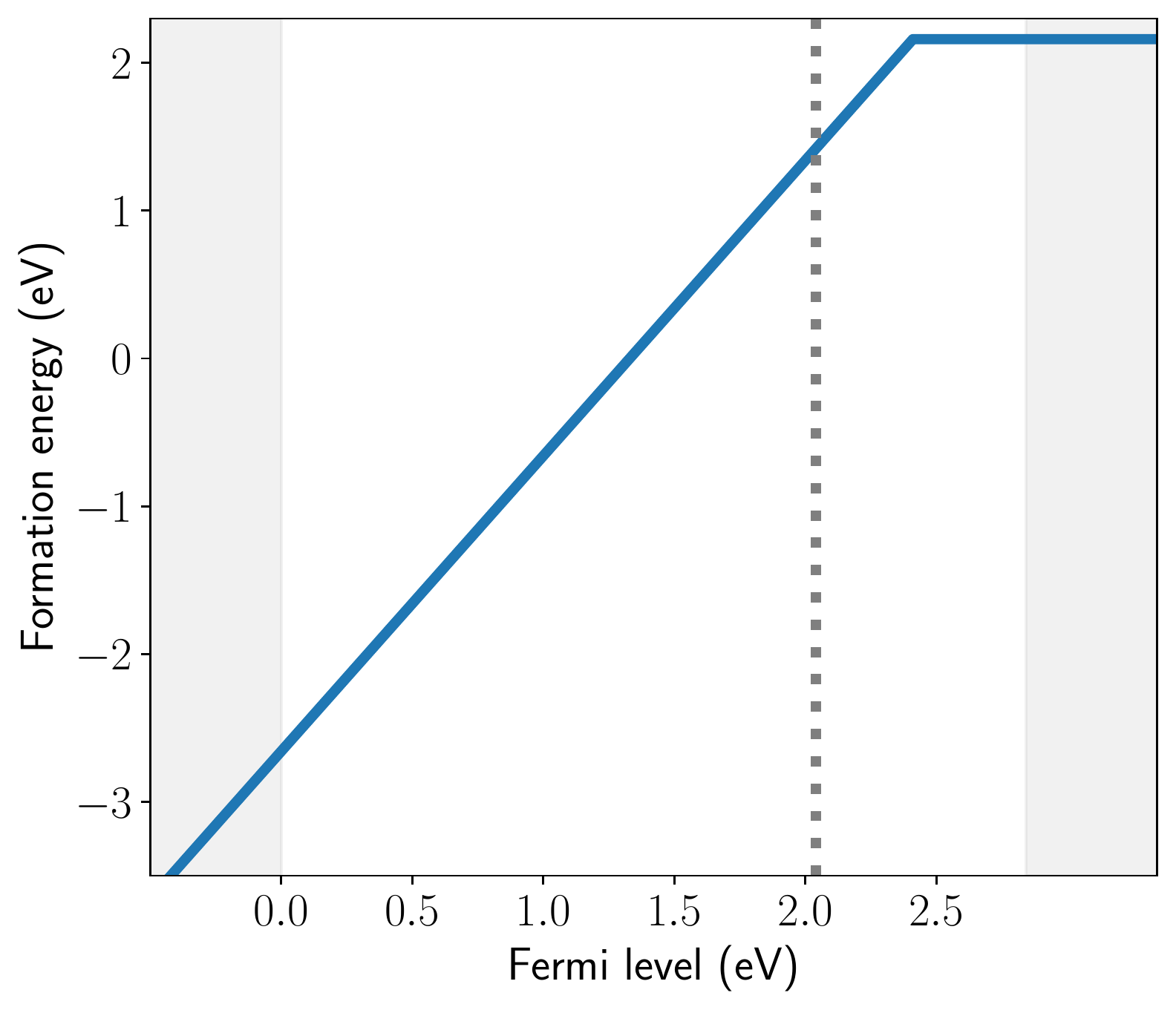}
    \caption{Formation energy for the surface oxygen vacancy in \ce{BiVO4}. The dashed vertical line indicates the \gls{ctl} of the electron polaron in defect-free \ce{BiVO_4}. The valence and conduction bands are shown in grey.}
    \label{fig:formen}
\end{figure}

Removing an oxygen atom from the surface of \ce{BiVO_4} leads to significant distortions of the lattice. The vanadium atom closest to the vacancy is pushed into the surface, sharing an oxygen atom with a subsurface \ce{VO_4} to retain its four-fold coordination. The introduction of the oxygen vacancy additionally frees up two excess electrons, which may localize at different sites in the lattice. In the stablest configuration, one electron localizes onto the corner-sharing V below the vacancy and the other at a nearby subsurface \ce{V} site. It may be energetically favorable for the vacancy to be ionized instead, however. We investigate this by computing the defect formation energies, shown in Fig. \ref{fig:formen}. Here, the chemical potential of oxygen is fixed to values corresponding to the oxygen-rich limit in our previous work.~\cite{osterbackaChargeLocalizationDefective2022} Also shown is the \gls{ctl} of the electron polaron in the subsurface, which is the preferred localization site for excess electrons in stoichiometric \ce{BiVO_4}. Only the (+2/0) \gls{ctl} lies within the bandgap, and the +1 charge state is thus thermodynamically unstable. The \gls{ctl} of the vacancy lies 0.4 eV above that of the polaron, however, indicating that it is energetically favorable for the excess electrons stemming from the vacancy to form small polarons in stoichiometric regions of the material rather than occupying states associated with the oxygen vacancies. In other words, the vacancy should spontaneously ionize and thus predominantly be in the ionized +2 charge state.

\begin{figure}
    \centering
    \includegraphics[height=400pt]{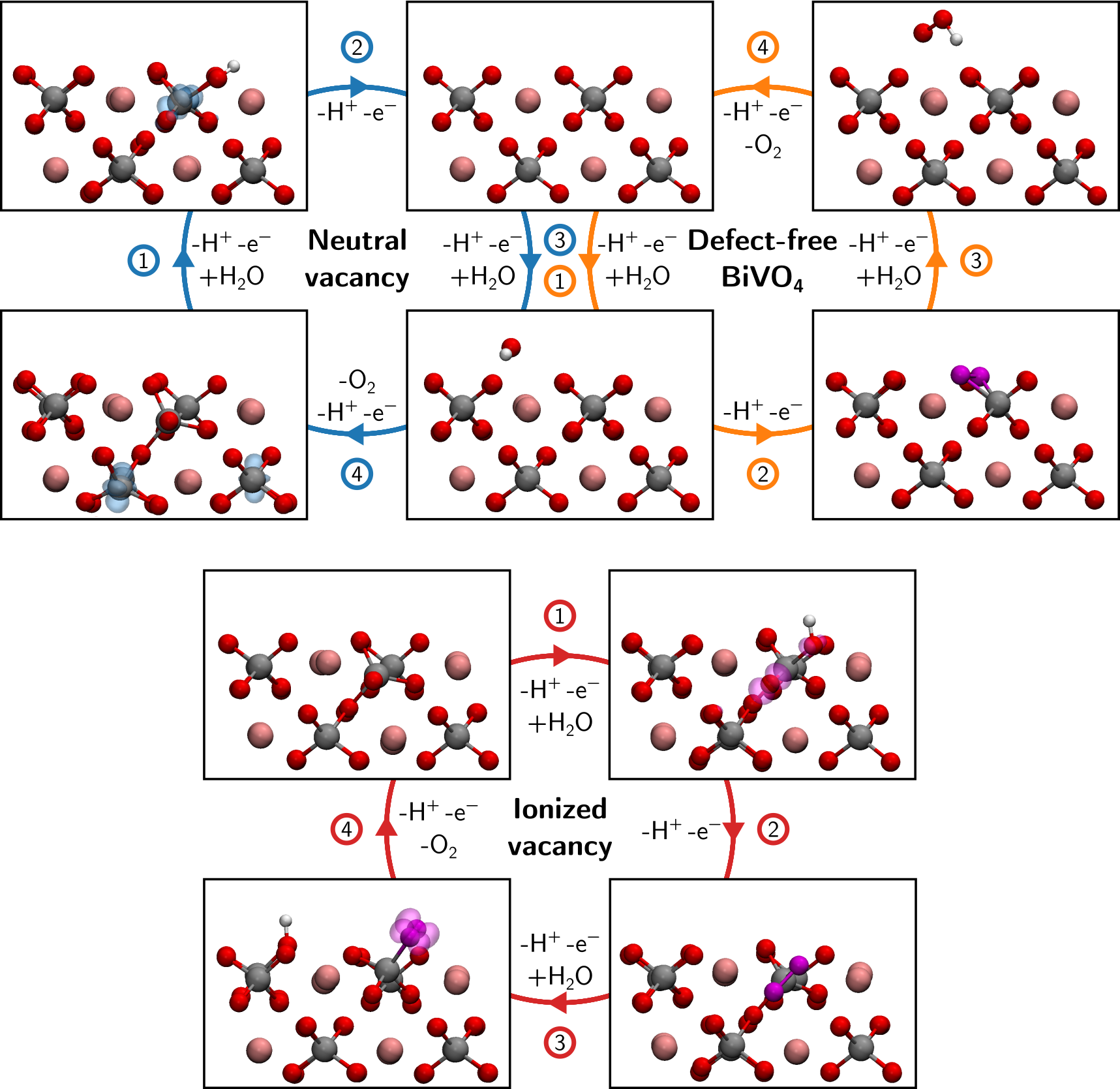}
    \caption{Schematic illustration of the \gls{oer} reaction mechanism showing side views of the surface structure of each intermediate. The inset numbers correspond to the reaction step numbering in Eq. \eqref{eq:watersplitting}. Isosurfaces correspond to localized charges, with electrons colored blue and holes magenta. Oxygen atoms involved in dimerization are colored purple.}
    \label{fig:schematic}
\end{figure}

\begin{figure}
    \centering
    \includegraphics[width=0.6\textwidth]{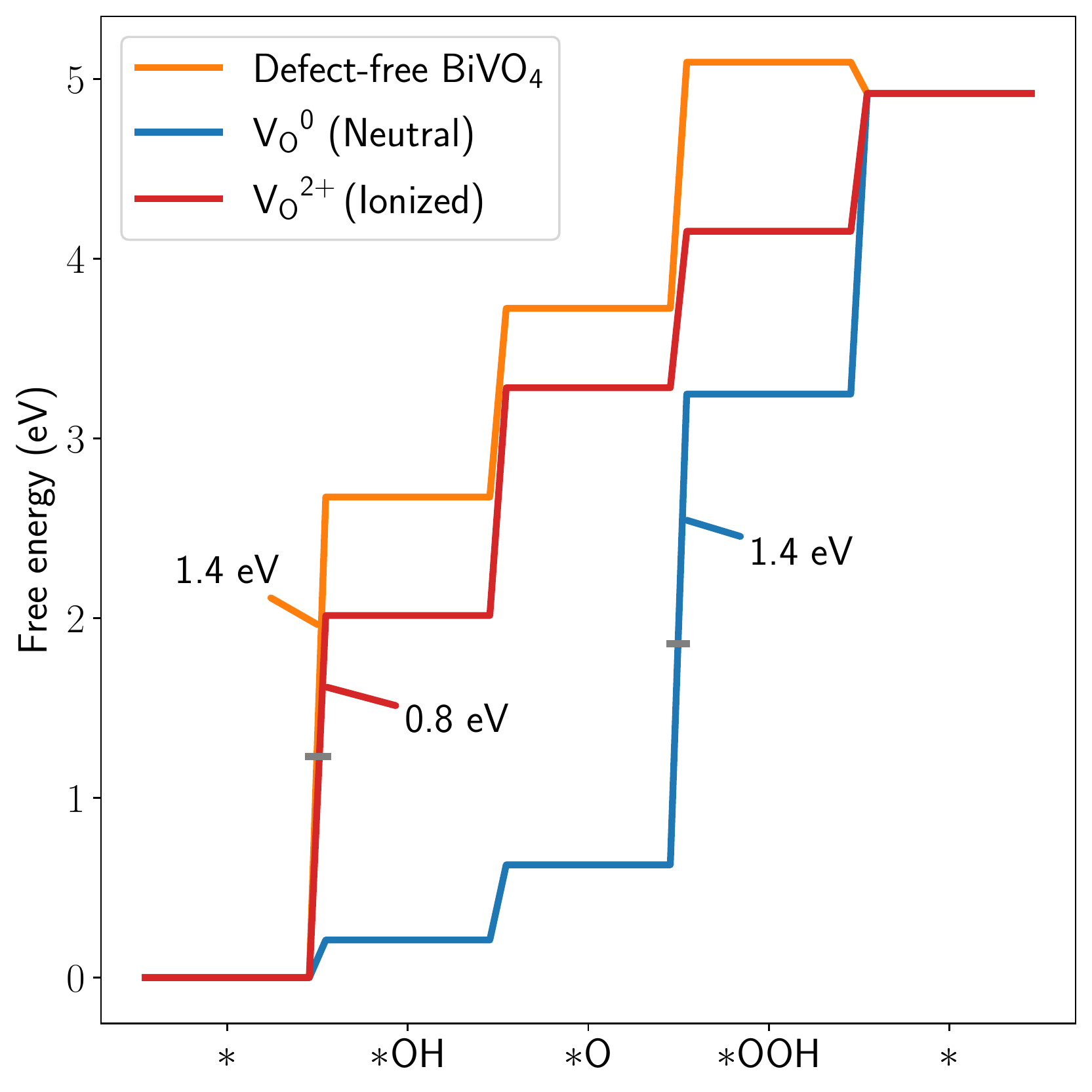}
    \caption{Free energy profiles for the \gls{oer} as shown in Eq. \ref{eq:watersplitting} on stoichiometric as well as oxygen-deficient \ce{BiVO_4}, with the latter shown in both neutral (0) and ionized (+2) charge states. The location of the ideal barrier height for each \gls{pds} is marked with a horizontal bar, and the excess barrier energy is also shown. These result in overpotentials of 1.4 V for the stoichiometric surface, 1.4 V for \ce{V_O^{0}}, and 0.8 V for the ionized \ce{V_O^{2+}}.}
    \label{fig:profiles}
\end{figure}

The effect of \ce{V_O} formation and ionization on the \gls{oer} mechanism is investigated by finding the stablest intermediates for the \gls{oer}, taking only the steps involving proton-coupled electron transfer into account. Renders of the resulting pathways for the reaction on the stoichiometric as well as the neutral (0) and ionized (2+) oxygen-deficient surfaces are shown in Fig. \ref{fig:schematic}. The free energy profiles of each reaction pathway, calculated within the \gls{che} formalism, are shown in Fig. \ref{fig:profiles}. This allows us to compare the efficiency of the mechanisms by computing their thermodynamic overpotentials. This is determined by the difference between the highest free energy difference of each reaction profile, i.e., the \gls{pds}, and the ideal free energy cost of 1.23 eV.

The primary active site on the stoichiometric surface is a bismuth atom, and the only difference to the reaction path proposed in literature is that the $\ce{$\ast{}$O}$ step involves the formation of an oxygen dimer with a bond length of 1.45 Å. The \gls{pds} is the very first step, ending up with an \ce{OH} group adsorbed onto a surface bismuth site, with an overpotential of 1.4 V. With the neutral vacancy, the reaction instead proceeds through filling the vacancy. The stoichiometric surface is recovered after the second reaction step, highlighted by the shared intermediates in \ref{fig:schematic}. The \gls{pds} is the third step, resulting in a \ce{Bi}-adsorbed \ce{OH} group and an overpotential of 1.4 V. In other words, the \gls{pds} is the same for both of these cases. The fourth step of the latter reaction reintroduces the \ce{V_O}, resulting in a higher free energy cost than the corresponding (third) step on the stoichiometric surface, involving only a proton-electron pair.

On the surface with \ce{V_O^{2+}}, the vacancy is the active site. The \gls{pds} is the first step, with an overpotential of 0.8 V, resulting in a \ce{OH} group filling the \ce{V_O}. As the positive charge state is fixed throughout the entire reaction, an excess hole is localized in close vicinity of the adsorbate. The second reaction step yields the stoichiometric structure, but with two excess holes. These localize onto two surface oxygen atoms, resulting in the formation of an oxygen dimer, or a peroxide ion (\ce{O_2^{2-}}), with a bond length of 1.45 Å. After the third reaction step, yet another excess hole localizes onto the adsorbed dimer, resulting in a surface-bound superoxide ion (\ce{O_2^{1-}}) with \ce{O-O} bond length 1.31 Å. After the removal of the final hydrogen atom, an oxygen molecule is evolved and the vacancy reintroduced at a significantly lower free energy cost than in the unionized case.

\mhchemoptions{arrow-min-length=1em}

\section{Discussion}

The formation energies for the surface \ce{V_O} as shown in Fig. \ref{fig:formen} show that the defect spontaneously ionizes no matter the Fermi level. It is always energetically favorable for the excess electrons stemming from the vacancy to localize in stoichiometric regions of the material, neglecting potential interactions with other defects. This is in agreement with experiments performed by Selim \textit{et al.}, the results of which suggest that surface oxygen vacancies in \ce{BiVO4} are found primarily in the ionized charge state in the dark.~\cite{selimImpactOxygenVacancy2019} We note that this stands in stark contrast to the computational results of Wang \textit{et al.}~\cite{wangRoleSurfaceOxygen2020} To compute formation energies in the dilute limit, finite-size correction schemes are used to account for spurious interactions between charged defects. This is typically done through dielectric continuum models, where the real material is approximated as a dielectric medium with embedded model charge distributions acting as stand-in for the defects. In the method proposed by Freysoldt and Neugebauer for asymmetric slabs,~\cite{freysoldtFirstprinciplesCalculationsCharged2018} a potential alignment step ensures that the model charges are placed in a reasonable manner. Wang \textit{et al.} modified this scheme for symmetric slabs, which renders potential alignment unnecessary. This introduces arbitrariness in how model charges are placed, which may contribute to the discrepancy we observe. However, our conclusions remain the same even when symmetric slabs are used with model charge placement taken from the asymmetric case, as shown in the SI.

The spontaneous ionization of surface \ce{V_O} into \ce{V_O^{2+}} additionally implies that they do not act as hole trapping sites unlike neutral \ce{V_O^0}, in agreement with the calculations of Cheng \textit{et al.}~\cite{chengControllingChargeCarrier2021} This increases the number of photogenerated holes that are able to participate in the \gls{oer}. Furthermore, the concentration of \ce{V_O} is not measured directly in experiments. Instead, the concentration of near-surface paramagnetic \ce{V^{4+}} is probed through electron spin resonance.~\cite{tanEnhancingPhotoactivityFaceted2017, selimImpactOxygenVacancy2019} These paramagnetic centers arise from the localization of excess electrons stemming from vacancy formation, reducing \ce{V^{5+}} in the surface region. The spontaneous ionization of surface \ce{V_O} implies that the elevation in \ce{V^{4+}} concentration associated with oxygen vacancy enrichment stems from subsurface polarons rather than states directly associated with the defects themselves. The energy landscape of intrinsic defects in \ce{BiVO4} is complex,\cite{zhangEffectsNativeDefects2020,osterbackaChargeLocalizationDefective2022} and disentangling oxygen vacancy-related \ce{V^{4+}} to those stemming from other intrinsic donor defects would be nigh impossible. The 2:1 \ce{V^{4+}-V_O} relationship assumed when reporting oxygen vacancy concentrations in literature may thus be flawed, though more detailed studies should be conducted to clarify this.

The optimal reaction pathways outlined in Fig. \ref{fig:schematic} for the stoichiometric as well as the neutral oxygen-deficient surfaces agree well with mechanisms previously considered in literature.~\cite{huAnisotropicElectronicCharacteristics2018,nikacevicInfluenceOxygenVacancies2021} The two mechanisms share two intermediates. Once they reach the step corresponding to OH adsorption onto the stoichiometric structure the reaction may proceed either through removal of proton-electron pair, or through the removal of both a proton-electron pair and an oxygen molecule. The latter process reintroduces the vacancy, and the pathway involving surface \ce{V_O^0} is thus equivalent to starting from the perfect structure and removing a lattice oxygen atom during one step of the reaction. \gls{oer} pathways that directly involve lattice oxygen have been proposed for several metal oxides,~\cite{binningerThermodynamicExplanationUniversal2015,meffordWaterElectrolysisLa12016,grimaudActivatingLatticeOxygen2017,yooRoleLatticeOxygen2018,zagalskayaInitioThermodynamicsKinetics2021} including bismuth vanadate,~\cite{hermansBiVO4SurfaceReduction2019} but our results suggest that the mechanism fails to increase the \gls{oer} activity in comparison to stoichiometric \ce{BiVO4}.

Nikačević \textit{et al.} propose an alternative reaction mechanism on the neutral oxygen-deficient surface, where the intermediates adsorb onto a surface bismuth atom next to the defect instead of filling the vacancy.~\cite{nikacevicInfluenceOxygenVacancies2021} While they found that this scheme results in a lower overpotential, they also found that its reaction intermediates are less stable than those of the vacancy-filling pathway. Based on energy differences, they estimate that around 97 \% of the surface vacancies would be filled at room temperature. This mechanism was therefore not considered in the present work as it is unlikely to be the primary driving force behind the \gls{oer} activity of \ce{BiVO4}.

On the surface with \ce{V_O^{2+}}, the \gls{oer} proceeds through filling and subsequent reintroduction of the vacancy in a fashion similar to the mechanism involving \ce{V_O^0}. This process is associated with a significantly lower energy cost for the ionized surface, however, which also exhibits a thermodynamic overpotential that is 0.6 V lower than the other cases considered here. The reaction thus becomes more favorable upon vacancy ionization. The mechanism we propose involves the formation of surface-bound peroxide and subsequently superoxide ions. Wang \textit{et al.} have proposed a similar mechanism for the \gls{oer} on \ce{TiO2},~\cite{wangIdentifyingKeyObstacle2018} and the surface-bound peroxide ion on \ce{BiVO4} has been predicted to be stable even in the presence of water by Wiktor and Ambrosio.~\cite{ambrosioStrongHoleTrapping2019} The last \gls{oer} step consists of the reaction {\ce{O_2^{1-} + h^+ -> O_2}}. This can be expected to occur quickly upon superoxide generation, as the standard reduction potential for the {\ce{O_{2}/O_2^{1-}}} redox couple lies around 0.33 V above the conduction band minimum of \ce{BiVO4},~\cite{monfortBismuthVanadatebasedSemiconductor2018} thus allowing for rapid hole injection to drive the reaction. As a consequence, the lifetime of the superoxide intermediate is likely short.

As highlighted by our results as well as by Todorova and Neugebauer,~\cite{todorovaIdentificationBulkOxide2015} the effect of defects on the electrochemical properties of a material depends on their charge state. For the \ce{V_O} in \ce{BiVO4} only the fully ionized defect is thermodynamically stable, so its beneficial effect should always be present. This need not be the case for other \gls{oer} photoanode materials. This opens up the possibility of tailoring photoanode properties through Fermi level engineering by carefully controlling defect and dopant concentrations as has previously been done for e.g. the NV center in diamond and \ce{BaZrO_3} multilayer structures.~\cite{muraiEngineeringFermiLevel2018, chavarriaFermiLevelEngineering2020}

\section{Conclusion}

In conclusion, we find that surface oxygen vacancies in \ce{BiVO4} spontaneously ionize, and therefore do not act as hole traps unlike unionized vacancies. Computational investigation of the oxygen evolution reaction intermediates show that this ionization enables a reaction pathway that exhibits a significantly lower thermodynamic overpotential than the optimal pathways on stoichiometric and unionized oxygen-deficient surfaces. Enriching the surface of \ce{BiVO4} thus makes the oxygen evolution reaction more favorable and improves the majority carrier concentration, while allowing for more photoexcited holes to partake in the reaction. The reaction intermediates in the pathway we propose on the ionized oxygen-deficient surface involves the formation of surface-bound perodixe (\ce{O_2^{2-}}) and superoxide (\ce{O_2^{1-}}) ions. We argue that the superoxide intermediate is very short-lived, allowing for rapid formation of oxygen molecules and the reintroduction of the vacancy to the surface. Our results shine new light on the nature of surface oxygen vacancies in \ce{BiVO4} and the manner in which they improve the photoelectrochemical activity as observed experimentally in oxygen-deficient samples of the material. They also highlight the importance of taking the full picture of defect and dopant occupancy into account when considering reactions involving non-stoichiometric surfaces.

\section{Acknowledgments}
The authors acknowledge funding from the ``Area of Advance - Materials Science'' at Chalmers University of Technology, and the Swedish Research Council (2019-03993). The computations were performed on resources provided by the National Academic Infrastructure for Supercomputing in Sweden (NAISS) at NSC and PDC.

\bibliography{mainrefs.bib}

\end{document}